\documentclass[%
 reprint,
 superscriptaddress,
 amsmath,amssymb,
 aps,
 prx,
 floatfix,
]{revtex4-2}

\usepackage{graphicx}
\usepackage{dcolumn}
\usepackage{bm}
\usepackage[normalem]{ulem} 
\usepackage{hyperref}

\usepackage{pifont}
\usepackage{braket}
\usepackage[usenames,dvipsnames]{xcolor}
\hypersetup{colorlinks=true, 
linkcolor=BrickRed, 
citecolor=Violet, 
filecolor=OliveGreen, 
urlcolor=RoyalBlue, 
filebordercolor={.8 .8 1}, 
urlbordercolor={.8 .8 0}
}%
\usepackage[caption=false]{subfig}
\makeatletter
\newcommand{\colorcaption}[2][]{%
  \begingroup%
  \renewcommand{\@caption@fignum@sep}{ (color online). }%
  \caption[#1]{#2}%
  \endgroup%
}
\makeatother

\renewcommand{\(}{\left(}
\renewcommand{\)}{\right)}
\renewcommand{\[}{\left[}
\renewcommand{\]}{\right]}
\renewcommand{\b}[1]{\mathbf{#1}} 

\newcommand{\T}{\mathcal{T}}

\DeclareMathOperator{\degree}{^{\circ}}

\begin{document}

\preprint{APS/123-QED}

\title{Competing magnetic states in transition metal dichalcogenide moir\'e materials}

\author{Nai Chao Hu}
 \affiliation{Department of Physics, The University of Texas at Austin,  Austin, TX 78712, USA.}

\author{Allan H. MacDonald}
\affiliation{Department of Physics, The University of Texas at Austin,  Austin, TX 78712, USA.}

\date{\today}

\begin{abstract}
Small-twist-angle transition metal dichalcogenide (TMD) heterobilayers develop 
isolated flat moiré bands that are approximately described by triangular lattice generalized 
Hubbard models \cite{fengchengHubbard}.  
In this article we explore the metallic and insulating states 
that appear under different control conditions
at a density of one-electron per moir\'e period, and the transitions between them.
By combining fully self-consistent Hartree-Fock theory 
calculations with strong-coupling expansions around
the atomic limit, we identify four different magnetic states and one nonmagnetic state
near the model phase diagram's metal-insulator phase-transition line.  
Ferromagnetic insulating states, stabilized by non-local direct exchange interactions,
are surprisingly prominent.
\end{abstract}

\maketitle


\section{\label{sec:intro}Introduction}

Moir\'e materials, formed by stacking layered 2-dimensional (2D) van der Waals 
semiconductors or semimetals with small differences in lattice constant or orientation, 
have attracted attention recently as a highly tunable platform to study strong correlation phenomena.  
The low energy physics of a moir\'e material is accurately described by an emergent periodic Hamiltonian \cite{Bistritzer12233,PhysRevB.82.121407,fengchengHubbard} 
that is insensitive to commensurability between the moir\'e pattern and the underlying lattice. 
Stimulated by the recent experimental realization \cite{kim2017tunable,cao2018correlated,cao2018unconventional} of magic angle physics in twisted bilayer graphene, expeimental attention has expanded to 
include other graphene based multilayers with twists   \cite{Yankowitz1059,lu2019superconductors,sharpe2019emergent,Serlin2020IntrinsicQA,cao2020tunable,shen2020correlated,chen2019evidence,chen2019signatures}, and also  
twisted transition metal dichalcogenide bilayers \cite{tang2020tTMD,regan2020mott,wang2020correlated,xu2020correlated,jin2021stripe,li2021charge,li2021imaging,fractionTMD2020,mak2021continuousMIT,dean2021quantumcritical,li2021quantum}.  
The valence bands of TMD heterobilayers 
and $\Gamma$-valley homobilayers \cite{mattia2020gammaValley} are 
described by emergent models in which interacting spin-$1/2$ electrons 
experience an external potential with triangular lattice periodicity, 
and therefore map directly to models of electrons on triangular or honeycomb lattices.  
This paper is devoted to a study of the properties of
triangular lattice moir\'e materials
and focuses on the case of one-hole per moir\'e period, where 
correlations are strongest.  
We examine the crossover from the narrow-band regime at 
small twist angles, 
where the system maps to a one-band Hubbard model with dominant on-site interactions,
to the regime closer to the metal-insulator phase transition where important differences appear.

\begin{figure}[t]
    \centering
    \includegraphics[width=0.45\textwidth]{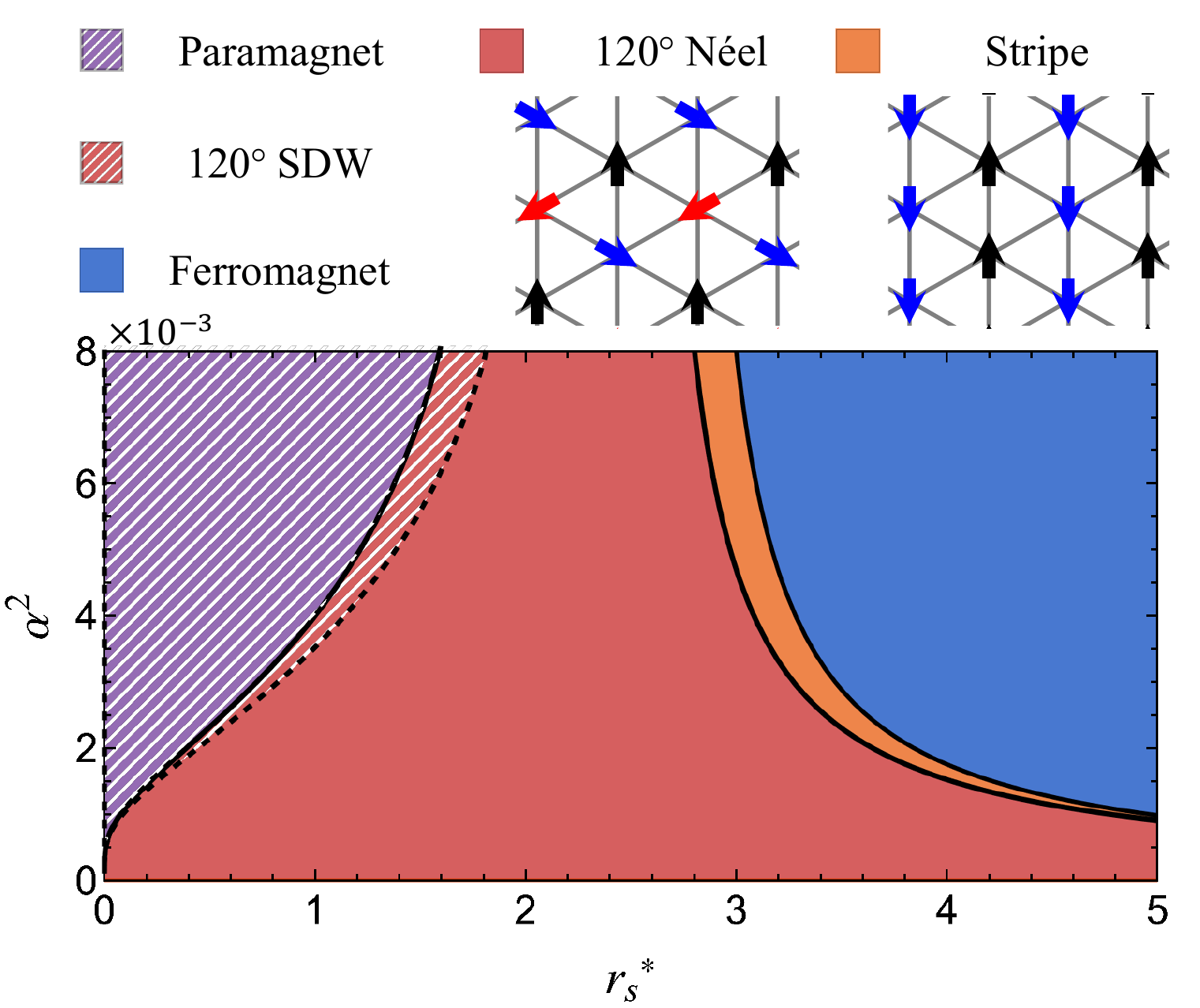}
    \caption{Hartree-Fock phase diagram for triangular lattice moir\'e materials 
    with one hole per unit cell. The two dimensionless control parameters (see main text)
    are the interaction strength $r_s^*$ and $\alpha^2$ - a parameter that 
    is inversely related to the moir\'e potential strength. 
    First order and second order phase transitions are marked by solid and dashed black lines,  respectively. States close to the top-left corner of the phase diagram (hashed) are metallic. States at the bottom right of the phase diagram are insulating.  A narrow semi-metallic state (labelled SDW - spin density wave) 
    that shares spatial symmetries with the three-sublattice non-collinear insulating state 
    hugs the metallic side of the metal-insulator transition. 
    An unexpected phase transition into an insulating ferromagnetic (blue) state at strong interaction strengths
    is interrupted by a narrow collinear antiferromagneric stripe state (orange). This phase diagram was calculated 
    for moir\'e modulation phase (see main text) $\phi=26\degree$.  The lines in this figure follow approximate phase boundary 
    expressions explained in the main text.
    }
    \label{fig:phaseD}
\end{figure}

Our discussion is based mainly on a mean-field Hartree-Fock approximation
used to address the interplay between periodic modulation
and Coulomb interactions that controls the hybridization 
between orbitals centered on different sites, and 
therefore exchange interactions of spins on the system's triangular lattice.  
Because it is a mean-field approach, the Hartree-Fock approximation
cannot account for dynamic fluctuations in spin-configuration,
but can accurately describe the energy of particular spin-configurations.  
Importantly for the present application, the Hartree-Fock approximation has the advantage over 
spin-density-functional theory \cite{liangfuCTI} 
that it correctly accounts for the absence of 
self-interaction \cite{selfInt81} when electrons are localized near lattice sites.
We expect the Hartree-Fock approximation to overestimate the stability of insulating states 
relative to metallic states. (Indeed this expectation is confirmed by comparisoin with separate exact-diagonalization calculations for the same model \cite{nicolas2020ed}.) 
Our calculations can therefore provide a lower bound on the 
moir\'e modulation strength that drives the system from a metallic to an insulating state
at a given interaction strength.  Unlike exact-diagonalization calcualations, Hartree-Fock calculations can be accurately converged with respect to system size.

Our goal in this manuscript is to identify differences between 
moir\'e material physics and single-band Hubbard model physics, with particular emphasis on the 
prospects for tuning the system into exotic spin liquid states. Fig.~\ref{fig:phaseD} shows the phase 
diagram in a space defined by dimensionless modulation strength $\alpha^2(V_M,\phi,a_M)$ and interaction 
strength $r_s^*(\epsilon, a_M)$ parameters. The full phase space of the problem is
actually 3-dimensional since the phase ($\phi$ - see below) of moir\'e potential Fourier amplitude also plays a role.
($a_M$ is the moir\'e material lattice constant.)
The lowest energy hole band is spectrally isolated for $\alpha\lesssim0.1$, the range covered in Fig.~\ref{fig:phaseD},
unless $\phi$ is very close to a honeycomb value.  (See below.)
We find that the three sublattice antiferromagnetism expected~ \cite{JolicoeurPhysRevB1990} in the insulating state 
transform to stripe magnetism and finally to ferromagnetism with increasing $r_s^*$, and that a semimetallic state 
with three sublattice order occurs on the metallic side of the metal-insulator phase transition.  
The transition to ferromagnetic insulating states at strong interactions 
opens up new opportunities to engineer strongly frustrated quantum magnetism.
Given the possiblity of {\em in situ} tuning between different spin states,
these findings demonstrate that moir\'e materials are an exceptionally promising new system for the 
exploration of two-dimensional quantum magnetism.

The rest of the paper is organized as follows: 
In Sec.~\ref{sec:hf} we review the moir\'e material model, discuss expected properties,
and introduce the mean-field formalism. In Sec.~\ref{sec:results}, 
we discuss our results for spin-interactions in insulating moir\'e materials. 
We comment specifically on necessary conditions for non-zero Hall conductance, 
concluding that though non-trivial band topology is unlikely, applying a magnetic 
field might induce a non-zero Hall conductance in doped insulators.  Finally in 
Sec.~\ref{sec:outlook} we summarize our results and highlight important directions for 
future research.  

\section{\label{sec:hf}Interacting continuum model}
The low energy physics of TMD moir\'e materials, like that of twisted bilayer graphene, is most conveniently captured by a continuum model \cite{fengchengHubbard}. Since TMDs are generally good insulators with strong spin-orbit coupling
from the transition metal atoms, only the topmost valence band needs to be included in the low energy model - yielding one 
state for each valley.  Because of spin-valley locking, we can equivalently choose to 
identify these states by their spins or by their valleys.  
The type-II band alignment of TMD heterojunctions means that 
only one layer is active at low energy.  
Hence the fermion field operators $\psi^{(\dag)}_{\alpha}(\b r)$ in this model carry only one label, 
representing locked spin/valley, while the effect of the other layer is integrated 
out, appearing only only as a contribution to the moir\'e potential \cite{fengchengDelta,fengchengHubbard},
$\Delta(\bm{r})$.
When the Fourier expansion of $\Delta$ is truncated at the first shell of 
moir\'e reciprocal lattice vectors $\mathbf{b}_j$,
\begin{equation}
\Delta(\mathbf{r}) = \sum_{j=1}^6 V_j \exp\[i\mathbf{b}_j\cdot\mathbf{r}\],
\end{equation}
where $V_j = V_M\exp\[(-1)^{j-1}i\phi\]$.  $V_j$ and $\phi$ are material-dependent parameters, with 
$V_M$ characterizing the moir\'e modulation strength and $\phi$ its shape.
The single-particle Hamiltonian of the moir\'e continuum model 
\begin{align}
    H_{0} = T + \Delta(\mathbf{r}),\label{eq:model}
\end{align}
where $T$ is the single-particle kinetic energy operator for electrons with effective mass $m^*$,
is spin-independent.  In a plane-wave representation
\begin{align}
    {H_{0}}_{\b{b},\b{b}'}(\b{k}) 
    = -\frac{\hbar^2}{2m^*}\(\b{k}+\b{b}\)^2 \, \delta_{\b{b},\b{b}'} + \sum_{j=1}^6 V_j\delta_{\b{b}_j,\b{b}-\b{b}'},
\end{align}
where momentum $\b{k}$ is in the first moir\'e Brillouin zone and the $\b{b}$'s are moir\'e reciprocal lattice vectors.

The key feature of this single-particle model, as pointed out in Ref.~\cite{fengchengHubbard}, 
is that the lowest energy hole band is isolated and has a bandwidth $W$ that 
decreases exponentially with moir\'e period $a_M$. 
One physical intuition for this behavior is based on the observation that the 
model can be 
approximated, in the large-$a_M$ limit, as a lattice of weakly coupled harmonically confined electrons. 
Ignoring the ``remote-band'' holes for the moment, we can readily see that the largest effect of 
Coulomb interactions is to impose an energy penalty $U$ on doubly occupied sites,
which is the essence of Hubbard model physics. But in contrast to simplest nearest-neighbor Hubbard model,
the ratio of second nearest-neighbor hopping to the nearest-neighbor hopping can be increased by decreasing the modulation strength, a property easily explained in the harmonic oscillator approximation, increasing the magnetic
frustration of insulating states.

In addition to allowing flexible tuning of $U/W$, the sublattice content of the hexagonal Bravais 
lattice on which the model sits can be controlled.  The symmetry of the moir\'e potential 
can be changed from that of a triangular lattice to that of a honeycomb lattice, with perfect 
honeycomb behavior achieved at $\phi=60^{\circ}$, $\phi=180^{\circ}$ and $\phi=-60^{\circ}$.
The applicable value of $\phi$ can be adjusted experimentally by 
choosing different TMD heterojunctions \cite{liangfuCTI, nicolas2020ed}. 
Over a finite range of $\phi$ near the honeycomb values, the moir\'e potential has local minima 
at the honeycomb lattice sites.  When $\phi$ is close to one of the honeycomb values,
the potential minima at the two honeycomb sublattices differ slightly in value,
allowing inversion-symmetry breaking sublattice-mass terms to be added to the 
Hamiltonian when expressed in terms of its tight-binding model limit.
Here we focus on triangular lattice Hubbard model Mott physics by 
restricting our attention to the case of one electron per triangular lattice unit cell.
At this density the second minimum plays a role only over narrow 
ranges of $\phi$ by increasing the spread of ground state Wannier wave function and 
slightly altering the competition between different states.

The many-body Hamiltonian in TMD moir\'e materials has three terms: the kinetic energy,
the moir\'e modulation potential, and the Coulomb interaction term.  It follows that the 
many-body physics depends, up to an energy scale, on $\phi$ and on  two dimensionless parameters.
We choose to describe the phase diagram in terms of the standard electron gas interaction strength 
parameter,
\begin{align}
r_s^* = \frac{1}{\sqrt{\pi n}a_B^*} = \(\frac{3}{4\pi^2}\)^{1/4}\frac{V_C^s}{T^s} = \(\frac{3}{4\pi^2}\)^{1/4} \; \frac{a_M}{a_B} \, \frac{m^*}{m}\, \frac{1}{\epsilon}, \label{eq:rs}   
\end{align}
and a second parameter that characterizes the ratio of the single-particle Wannier function spread to the 
moir\'e period.  Here have defined two energy scales: the kinetic energy scale at the 
moir\'e length $T^s=\hbar^2/2m^*a_M^2$ and the Coulomb interaction at the 
moir\'e length $V_C^s = e^2/2\epsilon a_M$, and one length scale - the Bohr radius 
$a_B^* \equiv \hbar^2\epsilon/e^2m^*$.  Our interaction strength parameter 
$r_s^*$ can be viewed as the typical distance 
between electrons in Bohr radius units.  The definition of the second dimensionless parameter 
is motivated by the small twist angle limit in 
which the lowest energy flat-band's Wannier function is accurately approximated by the 
Gaussian ground state of the harmonic potential $m^*\omega^2\b{r}^2/2$ obtained by 
expanding the moir\'e modulation potential around a minimum.  It follows from this expansion that 
$\omega^2 = \beta V_M/m^*a_M^2$ where  
$\beta = 16\pi^2\cos(\phi+k 120\degree)$, with $k$ an integer chosen to place the argument of 
the $\cos$-function $ \in (-60^{\circ},60^{\circ})$
\footnote{This choice of $\beta$ assures that the harmonic expansion is 
performed around the global minimum of the modulation potential.  The properties of TMD
moir\'e materials are invariant under $\phi \to -\phi$, and $\phi \to \phi + k 120^{\circ}$.}.
We choose 
\begin{align}
\alpha \equiv \frac{a_W^2}{a_M^2} = \frac{\hbar}{\sqrt{m^*\beta V_Ma_M^2}},
\end{align}
where $a_W^2 = \hbar/m^*\omega$ is the square of the oscillator length scale, 
as the second dimensionless interaction parameter.
Notice that $\alpha$ is dependent on both the phase $\phi$ and the magnitude $V_M$ of the moir\'e potential. 
In Fig.~\ref{fig:phaseD} the phase diagram is plotted in terms of $r_s^*$ and $\alpha^2$, with the latter 
variable chosen to simplify its dependence on $V_M$.  Choosing $\alpha$ as a dimensionless model 
parameter eliminates most of the phase diagram's dependence on $\phi$, with exceptions applying very 
close to $\phi\sim 60\degree + k 120\degree$, and deep in the metallic state.

\subsection{Symmetries}
For later convenience, we briefly summarize the symmetries of the problem. 
The model has a full SU(2) rotation symmetry of the locked spin/valley degree-of-freedom
and $C_{3v}$ orbital symmetry; the  $D_{3h}$ symmetry of a TMD monolayer is reduced by stacking.
The nontrivial operations of $C_{3v}$ are rotation by 
$2 \pi /3$,
\begin{align}
   C_3:  \psi_{\alpha}(\b{x}) \rightarrow \psi_{\alpha}(R_{2\pi/3}\b{x}),
\end{align}
and the mirror operation $M_y$,
\begin{align}
    M_y:  \psi_{\alpha}(x,y) \rightarrow \psi_{\alpha}(x,-y).
\end{align}
In the absence of a magnetic field, the model is invariant under time reversal, which switches the spins:
\begin{align}
    \T: \psi_{\alpha}(\b{x}) \rightarrow i(\sigma^y)_{\alpha\beta} \, \psi_{\beta}(\b{x}),
\end{align}
where $\sigma^y$ is the Pauli matrix acts in the spin space. We note that two spins are only interchanged by time reversal symmetry (TRS):
\begin{align}
    H^{\downarrow}_{\b{b},\b{b}'}(\b{k}) = H^{\uparrow*}_{-\b{b},-\b{b}'}(-\b{k}),
\end{align}
where we make use again of the discrete translational invariance of our moir\'e system.
At this level of approximation, each spin projected Hamiltonian itself also satisfies a spinless TRS property:
\begin{align}
    H^{\uparrow/\downarrow}_{\b{b},\b{b}'}(\b{k}) = H^{\uparrow/\downarrow*}_{-\b{b},-\b{b}'}(-\b{k}).
\end{align}
We also note that since inversion symmetry is broken in TMD monolayers, unlike in twisted bilayer graphene, Berry curvature is not required to vanish identically throughout the moir\'e Brillouin zone.

\subsection{Weak and Strong Modulation Limits\label{sec:lim}}
Before carrying out detailed self-consistent Hartree-Fock calculations, we provide some orientation
by discussing some simple limits in which electronic properties are well understood.
We first consider the weak modulation limit, where bandwidths are large and 
moir\'e bands overlap, which should give rise to behavior 
close to that of the 2D homogeneous electron gas (jellium) model. 
It is well known that Hartree-Fock approximation fails badly for magnetic properties 
by predicting that 2D jellium is ferromagnetic
above a small value of $r_s \simeq 2.01$, whereas quantum Monte Carlo 
calculations \cite{Tanatar1989GroundSO, Attaccalite2002CorrelationEA} shows
that the paramagnetic fluid remains 
stable up to a much larger interaction strength $r_s \sim 25.56$.
In the opposite strong modulation limit electrons occupy Wannier functions 
centered on potential minima.  At one electron per moir\'e period, strong on-site Coulomb 
interactions leave only spin-degrees of freedom at low energies.
Because electronic correlations are less subtle in this limit, Hartree-Fock approximation predicts 
magnetic states much more reliably, and we can fit ground state energies to 
determine the parameters of spin Hamiltonians. 
The only quadratic spin Hamiltonian that satisfies all the symmetry requirement of our 
model is the isotropic Heisenberg
model:
\begin{align}
    H_{spin} = J_1\sum_{\braket{i,j}}\mathbf{S}_i\cdot\mathbf{S}_j+J_2\sum_{\braket{\braket{i,j}}}\mathbf{S}_i\cdot\mathbf{S}_j\label{eq:spin},
\end{align}
where $\braket{i,j}$ and $\braket{\braket{i,j}}$ label nearest-neighbor and next-nearest-neighbor interactions.
In the case of $J_1>0$, the classical-spin triangular-lattice ground state 
is the 120$\degree$ N\'eel state for $J_2/J_1<1/8$ and a stripe state 
for $1/8<J_2/J_1<1$ \cite{JolicoeurPhysRevB1990}. 
It's widely believed that quantum fluctuations play a vital role in determining the phase near $J_2/J_1=1/8$, 
although extensive numerical efforts over the years \cite{kaneko2014gapless, SLPhysRevB.92.041105, SLPhysRevB.92.140403, SLPhysRevB.93.144411, SLPhysRevB.96.075116, SLPhysRevLett.120.207203}
have not reached a clear consensus on the nature of the potentially exotic phase. 
When the strong modulation Hamiltonian 
is approximated by a Hubbard model both interactions are antiferromagnetic \cite{allanPhysRevB.37.9753}.
with $J_2/J_1 \ll 1$, except possibly very close to the metal-insulator phase transition.
In the present system, however, we find that insulating states are ferromagnetic at 
large $r_s^*$ and interpret this property as evidence for beyond Hubbard-model physics.
The band topology of the ferromagnetic state obviously has zero Chern number since
spin-projected bands are then time-reversal invariant.  The topology of the 120$\degree$ N\'eel 
is less obvious since $\uparrow$ and $\downarrow$ band states are mixed by the non-collinear spin structure.
We nevertheless find that they are topologically trivial, as we show in 
Sec.~\ref{sec:results}. For extremely localized electrons ($\alpha\rightarrow0$), 
all magnetic configurations will become degenerate.

\subsection{Self-consistent Hartree-Fock approximation}

\begin{figure}[!t]
    \centering
    \includegraphics[width=0.45\textwidth]{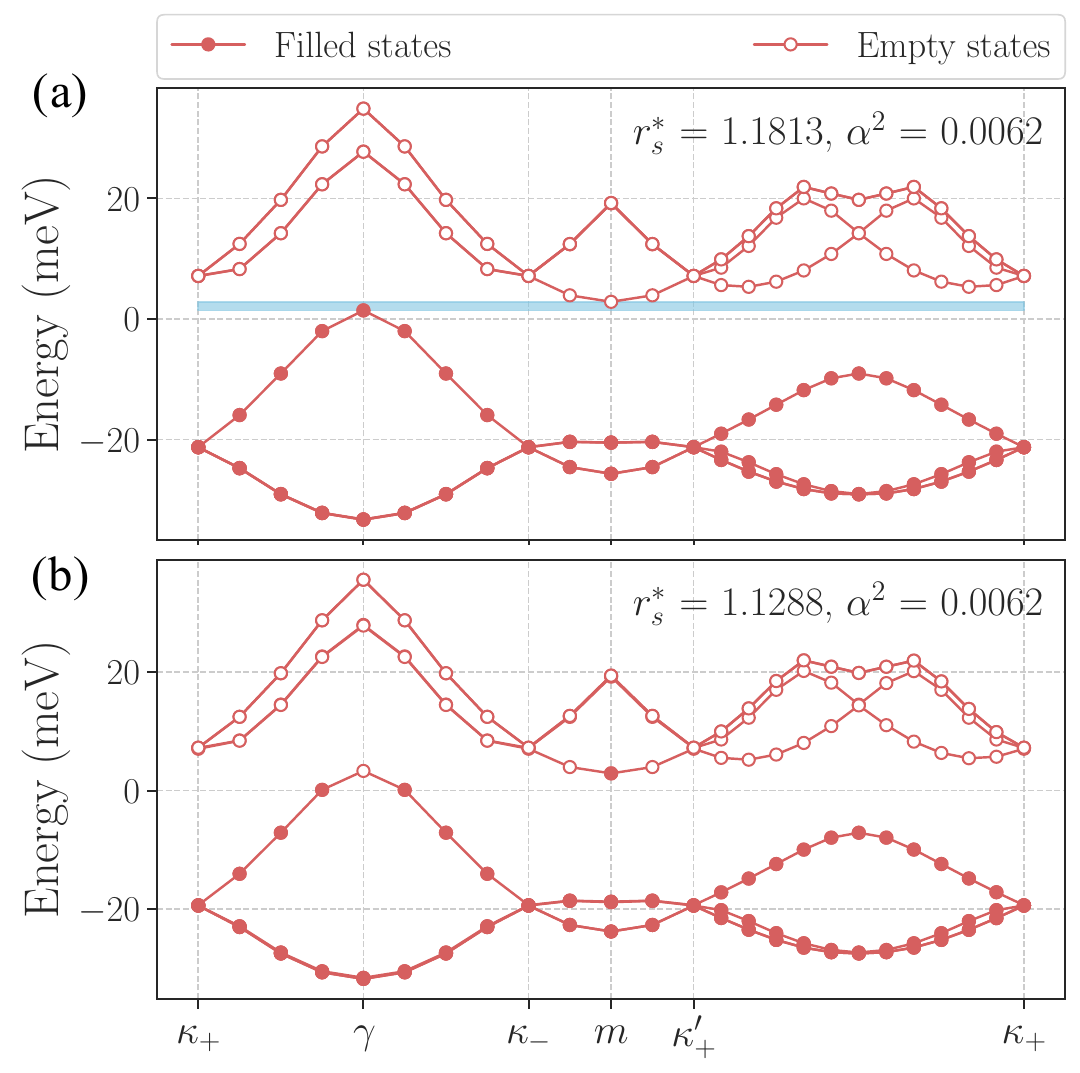}
    \caption{Typical Hartree-Fock band structures for three-sublattice magnetic states: (a) 120$\degree$ N\'eel insulator. The energy gap, indicated by the light-blue band, is small because the state is close to near the metal-insulator phase transition. 
    At smaller $r_s^*$, we obtain (b) a 120$\degree$ semimetallic SDW state. (See Appendix.~\ref{app:cutoff} for further 
    detail on model parameters.) 
    }
    \label{fig:hfbands}
\end{figure}

Our main goals in this MS are to gain insight into the spin-physics of the moir\'e superlattice
Mott insulator phase, 
and to obtain a rough estimate of the boundary between insulating and metallic states.
As long as the spin-ground state is close to its classical limit, that is to say as long as fluctuations in 
spin-direction in the magnetic ground state are not extremely large, the Hartree-Fock approximation is normally 
accurate.  One important advantage of the Hartree-Fock approximation is that it completely
removes spurious self-interaction effects in the limit that the electrons are reasonably
strongly localized around their moir\'e lattice sites.  As the twist angle is increased and the
moir\'e pattern's lattice constant is reduced, two-dimensional spin-density functional theory \cite{liangfuCTI},
which has much the same structure as Hartree-Fock theory, becomes an attractive alternative.
Even when quantum spin-fluctuations in the insulating ground state are large, either approach can 
be used to approximate the classical energy function of the spin subsystem, and quantum corrections 
can be calculated using standard spin-wave techniques.

The Hartree-Fock energy functional is the expectation value of the many-electron Hamiltonian in a 
single Slater determinant ground state.  Minimizing the energy functional with respect to 
single-particle wave-functions yields a mean-field Hamiltonian that adds an interaction self-energy 
$\Sigma^{HF}$ to the single-particle Hamiltonian which can be expressed 
in terms of the single-particle density matrix $\rho = \sum_{n} \ket{\psi_n}\bra{\psi_n}$, where the sum is 
over occupied moir\'e-band Bloch wavefunctions.  The mean-field electronic structure of moir\'e superlattices is best 
evaluated using a plane-wave representation in which the Hartree-Fock self energy $\Sigma^{HF}$ at each 
$\b{k}$ in the Brillouin-zone is a matrix in reciprocal lattice vectors $\b{b}$:
\begin{widetext}
\begin{equation}
    \Sigma^{HF}_{\alpha,\b{b};\beta,\b{b}'}(\b{k}) = \frac{\delta_{\alpha,\beta}}{A}\sum_{\alpha'}V_{\alpha'\alpha}(\b{b}'-\b{b})\sum_{\b{k}',\b{b}''}\rho_{\alpha',\b{b}+\b{b}'';\alpha',\b{b}'+\b{b}''}(\b{k}')  -\frac{1}{A}\sum_{\b{b}'',k'}V_{\alpha\beta}(\b{b}''+\b{k}'-\b{k})\rho_{\alpha,\b{b}+\b{b}'';\beta,\b{b}'+\b{b}''}(\b{k}').
\label{eq:self-energy}
\end{equation}
\end{widetext}
In Eq.~\ref{eq:self-energy} Greek letters label spin, $A$ is the finite sample area corresponding to a 
discrete Brillouin-zone mesh, and $\rho_{\alpha,\b{b};\beta,\b{b}'}$ is the self-consistently determined 
momentum-space density matrix.   Starting with a physically plausible density matrix $\rho_0$, we minimize the energy by performing self-consistent iterations. Because the many-body interaction is invariant under both translations and 
spin-rotations, if we start from a density matrix $\rho_0$ which satisfies a symmetry $\hat{O}$ of $H_0$ 
($[\rho_0,\hat{O}] = 0$, $[H_0,\hat{O}] = 0$) then the symmetry survives under iteration.  
That is to say that $H^{HF}$ commutes with $\hat{O}$ at every iteration step.  In many-cases the minimum 
energy Hartree-Fock state breaks symmetries of $H_0$ and these solutions are found under iteration only 
by starting with a broken-symmetry density-matrix. 
As argued in Sec.~\ref{sec:lim}, the phase diagram contains paramagnetic states that do not break 
any symmetries, ferromagnetic states with spontaneous 
collinear spin-polarization that do not break lattice translational symmetries,
stripe states with collinear order and a doubled unit cell area, 
and 120$\degree$ N\'eel states with both a tripled unit cell area and and non-collinear spin-order.
We obtain solutions of the first two kinds by appropriate choices of the initial density $\rho_0$.
Each possible type of reduced translational symmetry implies a different reciprocal lattice,
and therefore has to be encoded explicitly in the recriprocal lattice employed and considered separately.
Solutions can be classified as insulating with a gap between occupied and empty states, or 
metallic with Fermi surfaces in the Brillouin-zone on which occupation numbers change.  
At one electron per moir\'e period, the paramagnetic state must be metallic, but all other 
states we consider can be insulating. We show typical Hartree-Fock self-consistent band structures for insulating and metallic magnetic ordered sates in Fig.~\ref{fig:hfbands}.

\section{\label{sec:results}Results}
Having introduced the problem we now present the predictions of Hartree-Fock theory for the phase diagram. 
We focus first on a fixed moir\'e modulation phase $\phi = 26\degree$, estimated \cite{fengchengHubbard}
to apply to the WSe$_2$/MoSe$_2$ heterobilayer system.  
(As emphasized earlier we have chosen the dimensionless parameters used to construct the phase 
diagrams with a view toward minimizing any dependence on $\phi$.). We have performed self-consistent 
Hartree-Fock calculations on a discrete two-dimensional grid of system parameters.
The phase diagram in Fig.~\ref{fig:phaseD} was constructed by interpolating between this discrete set of 
results; a pixelated summary of our actual calculation results is presented in Fig.~\ref{fig:pixel}
of the supplementary material.  The influence of $\phi$ on the phase diagram will be discussed later.
 
Each solution of the Hartree-Fock equations corresponds either to a local minimum of the 
energy functional, or to a saddle point at which energy can be reduced by breaking symmetries.
We have identified the ground state by comparing the total energies of all solutions.
Translational symmetry is allowed to break down only to either the two sublattice (stripe) state or 
the three-sublattice (120$\degree$) state, both of which are common in triangular lattice phase diagrams.
In our calculations, all phase transitions that change translational symmetry are
of the first order, and all that do not are continuous.

\begin{figure}
\centering
\includegraphics[width=0.45\textwidth]{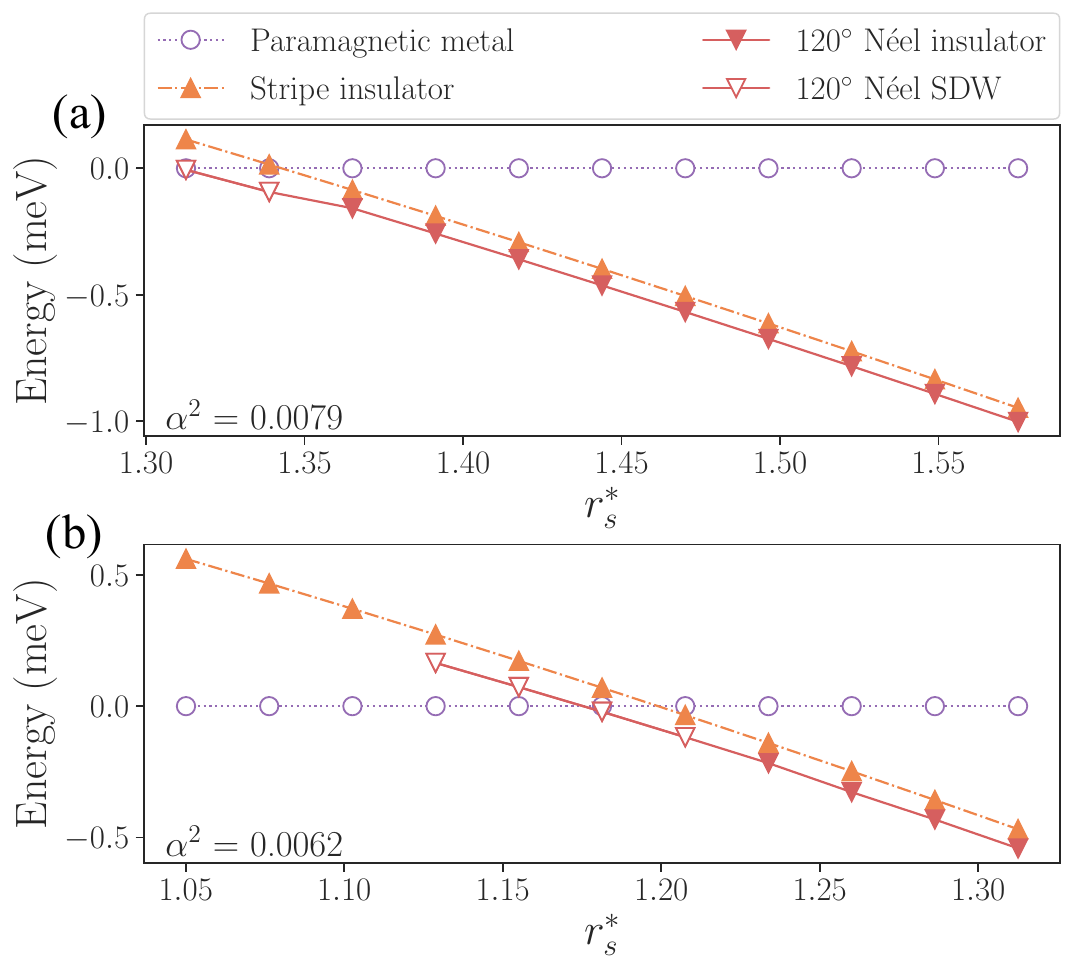}
\caption{We choose two values of the Wannier localization parameter $\alpha^2$ to 
closely examine the emergence of the 120$\degree$ semimetallic SDW state and the disappearance of metallic states with  
increasing $r_s^*$. Energies are plotted relative to the nonmagnetic metallic state energy. 
The ferromagnetic state has higher energy than both antiferromagnetic structures and is hence omitted in these plots. 
At the relatively large $\alpha^2$ values we choose, 
the lowest energy state clearly change from paramagnetic metal to a 120$\degree$ semimetallic SDW, and then to a 120$\degree$ antiferromagnetic insulator.}
\label{fig:mit}
\end{figure}

Two sets of two phase boundaries are of particular interest: metal-insulator transitions on the left-hand side of 
Fig.~\ref{fig:phaseD} and magnetic transitions within the insulating state on the right-hand side.
We see in Fig.~\ref{fig:phaseD} that the competition near the metal-insulator phase transition
is mostly between a non-magnetic metallic state and the noncollinear three-sublattice 
state.  When we examine the region near the metal-insulator phase transition closely, however,
we find that the three-sublattice insulator becomes a semimetal at a critical $r_s^*$ that is slightly larger than the critical $r_s^*$ 
at which the magnetic order disappears. (See Fig.~\ref{fig:mit}.)
As a result, itinerant magnets with the same magnetic structure as that of the non-collinear three-sublattice insulating state
appears near the metal-insulator transition.  
Thus within the Hartree-Fock approximation, the insulator-to-metal transition is a continuous phase transition, but is closely followed by a first order transition to 
a non-magnetic metallic state.  
We associate the increasing stability of 120$\degree$ semimetallic SDW states, relative to 120$\degree$ insulating 
states at larger values of $\alpha$ with increased increasing itineracy and associated larger values of  
$t_2/t_1$.  We also find that close to the metal-insulator phase boundary, the stripe and 120$\degree$ insulating states have 
very similar energy densities, although the stripe order energies are always slightly larger.

To gain some analytic insight into the form of this phase boundary, we make an approximation that is accurate 
in the small twist angle limit discussed earlier.   We estimate the nearest neighbor hopping parameter by using harmonic oscillator wave functions
which yields
\begin{align}
    t_1 = \frac{\hbar^2}{2m^*a_M^2}\(\frac{1}{4\alpha^2}-\frac{1}{\alpha}\)\exp\(-\frac{1}{4\alpha}\).\label{eq:mit}
\end{align}
The non-monotonic dependence of $t_1$ on $\alpha $ is related to a breakdown of the assumption of strongly localized 
Wannier orbitals at small $\alpha$.  To simplify the approximate phase boundary expression we derive below 
we measure energies in units of the moir\'e kinetic energy scale $T^s$ and write $\tilde{X} \equiv X/T^s$. 
On physical grounds, the metal-insulator transition should occur at a critical value of the ratio $c = U/t_1$. 
Since the ratio of the moir\'e lattice constant to the Wannier function width, which is $\propto \theta^{1/2}$ 
changes slowly in the parameter range of interest, this criterion corresponds 
approximately to a critical $c'$ of the ratio $V_C^s/t_1$, which is proportional to $r_s^*$. (See Eq.~\eqref{eq:rs}.) 
In estimating the phase boundary line we ignore the $1/\alpha$ factor in Eq.~\eqref{eq:mit} since $\alpha$ is small. 
This yields $\alpha^2 = [8W_{-1}(-\sqrt{\tilde{t}_1}/4)]^{-2}$, where $W_{-1}(x)$ is the Lambert $W$ function and the branch is chosen by the monotonic property of $t_1$.  The Hartree-Fock metal-insulator phase boundary closely follows the $c' = 1.9$ line in the phase diagram, 
which corresponds to $c=2c'/\sqrt{\alpha}\sim15.1$. 
Given this value for $c$, we can estimate that the magnetic ordering energy 
on the insulating side of the phase diagram $4t_1^2/U$ is $\sim 4.0 meV$ for 
experimental systems with moir\'e period $a_M\sim 5nm$ \cite{mak2021continuousMIT,dean2021quantumcritical}, 
which compares well to the experimental estimate of $J_1\sim3meV$ \cite{mak2021continuousMIT}. 

\begin{figure}
\centering
\includegraphics[width=0.45\textwidth]{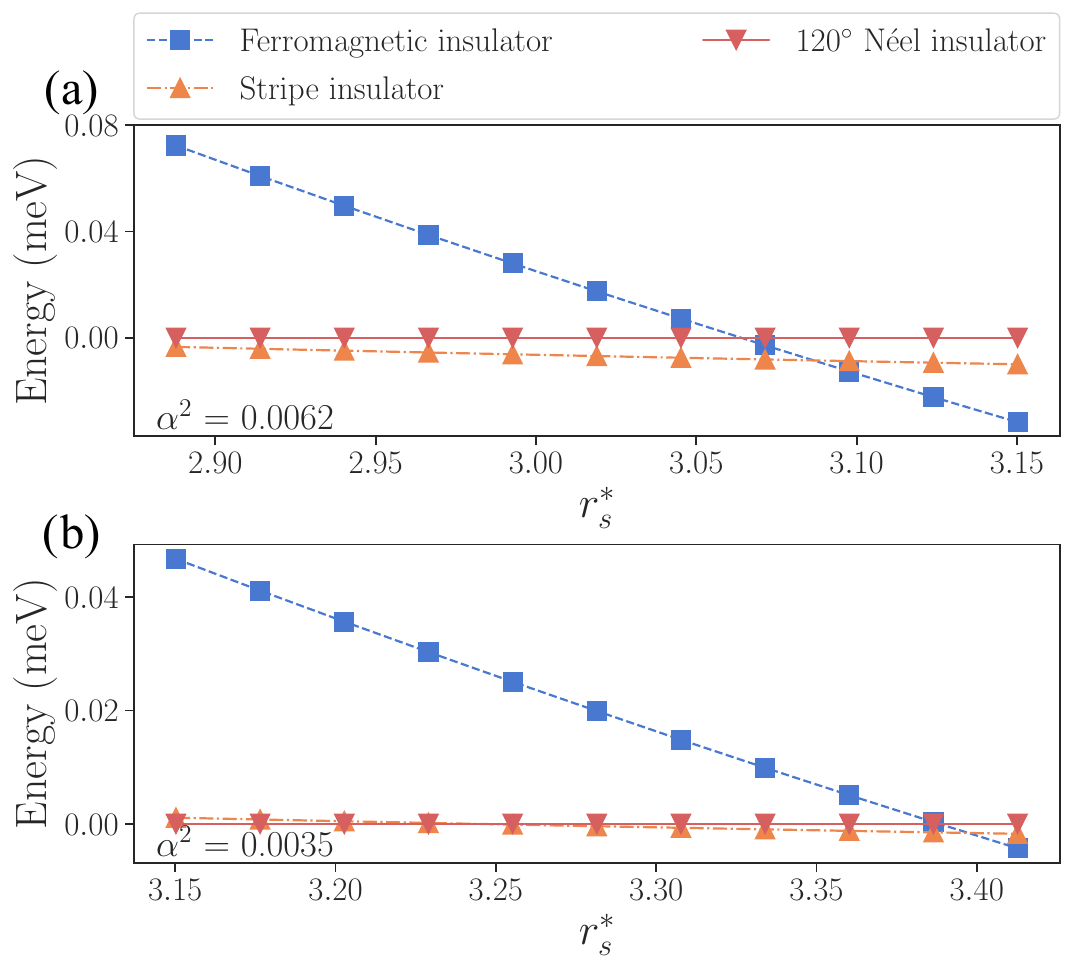}
\caption{Phase competition near the antiferromagnet-ferromagnet transition. Again the Wannier localization parameter $\alpha^2$ is fixed and interaction strength $r_s^*$ is varied. Energies are plotted relative to the 120$\degree$ N\'eel insulators. Paramagnetic metals in this case have much higher energy than any insulators and therefore are omitted. The lowest energy state changes from a 120$\degree$ antiferromagnetic insulator at small $r_s^*$ to a stripe insulator, and then to a ferromagnetic insulator at large $r_s^*$. The stripe phase becomes more stable at larger $\alpha$.}
\label{fig:fm}
\end{figure}

We now turn to the magnetic transitions that occur within the insulating
region in the phase diagram.  We find that the insulators are ferromagnetic at 
large $r_s^*$, and that all magnetic states are very close in energy near the 120$\degree$ state to 
ferromagnet phase boundary.  Evidently this phase boundary is associated with a change in sign of the dominant 
near-neighbor interactions between spins, leaving all states close in energy.
The energies of competing states close to this phase boundary are plotted in 
Fig.~\ref{fig:fm}, where we see that the ferromagnetic state is strongly favored 
at large interaction strength $r_s^*$, that stripe phase are stable over a narrow 
range of $r_s^*$ between the 120$\degree$ and ferromagnetic states, and that the 
stripe state stability range broadens at larger $\alpha$.

\begin{figure}
    \centering
    \includegraphics[width=0.45\textwidth]{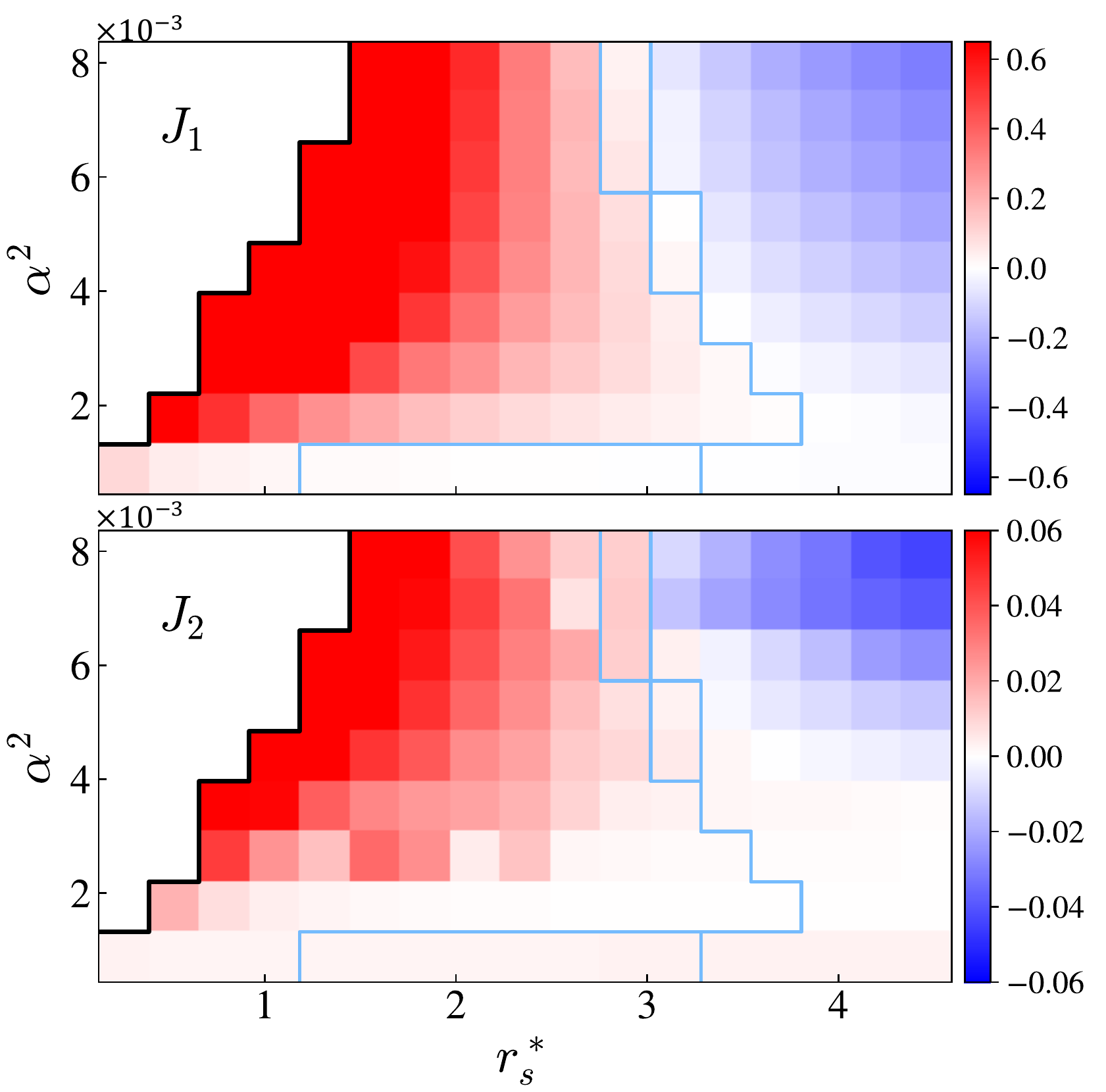}
    \caption{Values of $J_1$ and $J_2$ in units of $meV$ for a fixed $a_M$ throughout the phase diagram, obtained by fitting ground state energies of different insulating magnetic order to a $J_1$-$J_2$ Heisenberg model. (See Eq.~\eqref{eq:j1}-\eqref{eq:j2}.) We observe that $J_1$ changes sign at the antiferromagnet-ferromagnet phase boundary, while $J_2$ changes sign inside the region of ferromagnetic states.}
    \label{fig:js}
\end{figure}

To understand these observations, we consider interactions within the spin-only Hilbert space discussed in Sec.~\ref{sec:lim}. 
Assuming the spin Hilbert space is correctly described by the $J_1-J_2$ spin model of Eq.~\eqref{eq:spin}, 
the classical energies of the ferromagnetic, stripe, and 120$\degree$ states are 
\begin{align}
    e_{1} =& \frac{1}{4N}(3NJ_1+3NJ_2) = \frac{3}{4}J_1 + \frac{3}{4}J_2,\\
    e_2 =& \frac{1}{4N}(-NJ_1-NJ_2) = -\frac{1}{4}J_1 - \frac{1}{4}J_2,\\
    e_{3} =& \frac{1}{4N}\(-3NJ_1/2+3NJ_2\) = -\frac{3}{8}J_1 + \frac{3}{4}J_2.
\end{align}
It follows that can determine numerical values for the coupling constants from the energy differences between the 
three magnetic states we consider in our Hartree-Fock calculations:
\begin{align}
    J_1 =& \frac{8}{9}\(e_{1}-e_{3}\),\label{eq:j1}\\
    J_2 =& e_{1}-e_{2}-J_1. \label{eq:j2}
\end{align}
We plot the $J_1$ and $J_2$ values obtained in this way
in Fig.~\ref{fig:js}, where we see that the signs of $J_1$ and $J_2$ are strongly correlated, 
and that the ferromagnetic state phase boundary aligns with the line on which $J_1$ changes sign.

One of the most intriguing aspects of our results is the appearance (at the mean field level) of a stripe state.
This finding suggests that these moir\'e materials may provide a clean realization of the long-sought $J_1$-$J_2$ quantum spin liquid state, 
which is born out of the quantum fluctuations near the boundary between the three-sublattice state and the stripe state. 
However, we caution the readers that our system is not fully equivalent to a $J_1$-$J_2$ model. 
In the case of an exact $J_1$-$J_2$ Heisenberg model, stripe states appear for $J_2/J_1\gtrsim 1/8$. 
Since we estimate values of the exchange couplings by comparing energies of a small number of magnetic configurations, 
our results do not rule out other possibilities, one example of which is that the third nearest neighbour exchange coupling $J_3$ is ferromagnetic and $-J_3/J_1\gtrsim1/9$. Since current experimental studies operate in parameter ranges 
close to the metal-insulator transition, they may need to tune to larger $r_s^*$ to reach the ferromagnetic state, 
for example by choosing materials with smaller lattice mismatches or tuning twist angles.

The $J_1$ sign change is associated with interactions that are non-local in the model's 
Wannier function lattice representation \cite{[{See also }]nicolas2021}, and 
therefore absent in generalized Hubbard-model interaction approximations.
For a given pair of near-neighbor sites the non-local interaction terms can be 
characterized as either an interaction-assisted hopping term,
\begin{align}
    V_{ah} = \sum_{\sigma}\braket{2\sigma,1\bar{\sigma}|V_C|1\sigma,1\bar{\sigma}}c^{\dag}_{2\sigma}c^{\dag}_{1\bar{\sigma}}c_{1\bar{\sigma}}c_{1\sigma},
\end{align}
or as a intersite-exchange term,
\begin{align}
    V_{x} = \sum_{\sigma_1,\sigma_2}\braket{2\sigma_1,1\sigma_2|V_C|1\sigma_1,2\sigma_2}c^{\dag}_{2\sigma_1}c^{\dag}_{1\sigma_2}c_{2\sigma_2}c_{1\sigma_1},
\end{align}
where $\sigma$ is a spin label and $\bar{\sigma}=-\sigma$. At half-filling, $V_{ah}$ is physically equivalent to hopping, so its main effect is to enhance the 
antiferromagnetic coupling constant. On the other hand, $V_{x}$, being an exchange term, carries a minus sign from fermionic ordering and therefore favors ferromagnetic coupling \footnote{See  Appendix~\ref{app:tU} for a more formal explanation via an expansion around the flat-band limit.}.
We now argue the transition happens when the enhanced antiferromagnetic coupling $4(-t_1+|V_{ah}|)^2/U$ is equal to $2|V_x|$ in magnitude. 
We assume we are in the strongly interacting regime where $V_{ah/x}\sim t_1^2/U\ll t_1$. This allows us to compare the simplified antiferromagnetic energy scale $4t_1^2/U$ to ferromagnetic energy scale
\begin{align}
    2|V_{x} ({\b 1},{\b 2};{\b 2},{\b 1})| = 2U\exp\(-\frac{1}{2\alpha}\),\label{eq:x}
\end{align}
where $\b 1$ and $\b 2$ are nearest neighbors. We again ignore the $1/\alpha$ factor in the expression Eq.~\eqref{eq:mit} of $t_1$ to the lowest order. It's clear then the antiferromagnet-ferromagnet phase boundary should be described by $1/2\sqrt{2}\alpha^2 = \tilde{U} $, i.e. $\alpha^2 = 1/2\sqrt{2}\tilde{U} \sim 1/r_s^*$ \footnote{Increasing $\alpha^2$, the parameter that measures the relative spread of the Wannier functions,
increases both $t_1$ and $|V_x|$.  From Eq.~\eqref{eq:mit} and \eqref{eq:x}) we see that $|V_x|$ increases 
more rapidly than $t_1^2$.  This explains why $\alpha^2 \sim 1/r_s^*$ along the ferromagnetic/antiferromagnetic phase boundary.}, which agrees well with our Hartree-Fock approximation.

To explicitly explore the influence of $\phi$ on the phase boundaries that remains for our choice of 
dimensionless interaction parameters, we carry out self-consistent Hartree-Fock calculations {\it vs.} $\phi$ at two points in our $r_s^*-\alpha^2$ phase diagram (Fig.~\ref{fig:phaseD}) that lie just to the right of the two phase transition boundaries.  In order to describe how the 
generic triangular lattice smoothly evolves into a honeycomb lattice we consider the range from $\phi = 30\degree$ to 
$\phi=60\degree$ at which the additional honeycomb lattice symmetries become exact. 
Because we study the case of one-electron per-triangular lattice unit cell,
the electron density is half of that associated with honeycomb lattice Mott insulator states.
As $\phi$ approaches $60 \degree$, the two-local potential minima in the moiré unit call become more nearly equivalent, and inversion 
symmetry relative to the mid-point between the two minima is more nearly established.  
In our self-consistent Hartree-Fock calculations we find that at the density we study this approximate symmetry is always strongly 
broken.  Even at $\phi=60\degree$, the electrons tend to occupy one honeycomb sublattice only, as we verify by explicit caclulation,
and the role of the 
difference between $\phi$ and $60 \degree$ acts as a weak symmetry-breaking parameter.  At no point in 
this evolution do the lowest two self-consistent Hartree-Fock bands overlap and develop the Dirac points of single-orbital 
honeycomb lattice bands.  The broken symmetry lowers energies by increasing separations between electrons.
In the language of the honeycomb lattice Hubbard model, occupying only one honeycomb sublattice
avoids the near-neighbor electron-electron interaction term in the Hamiltonian with coupling constant 
$U_1$. 

\begin{figure}[t]
    \centering
    \includegraphics[width=0.45\textwidth]{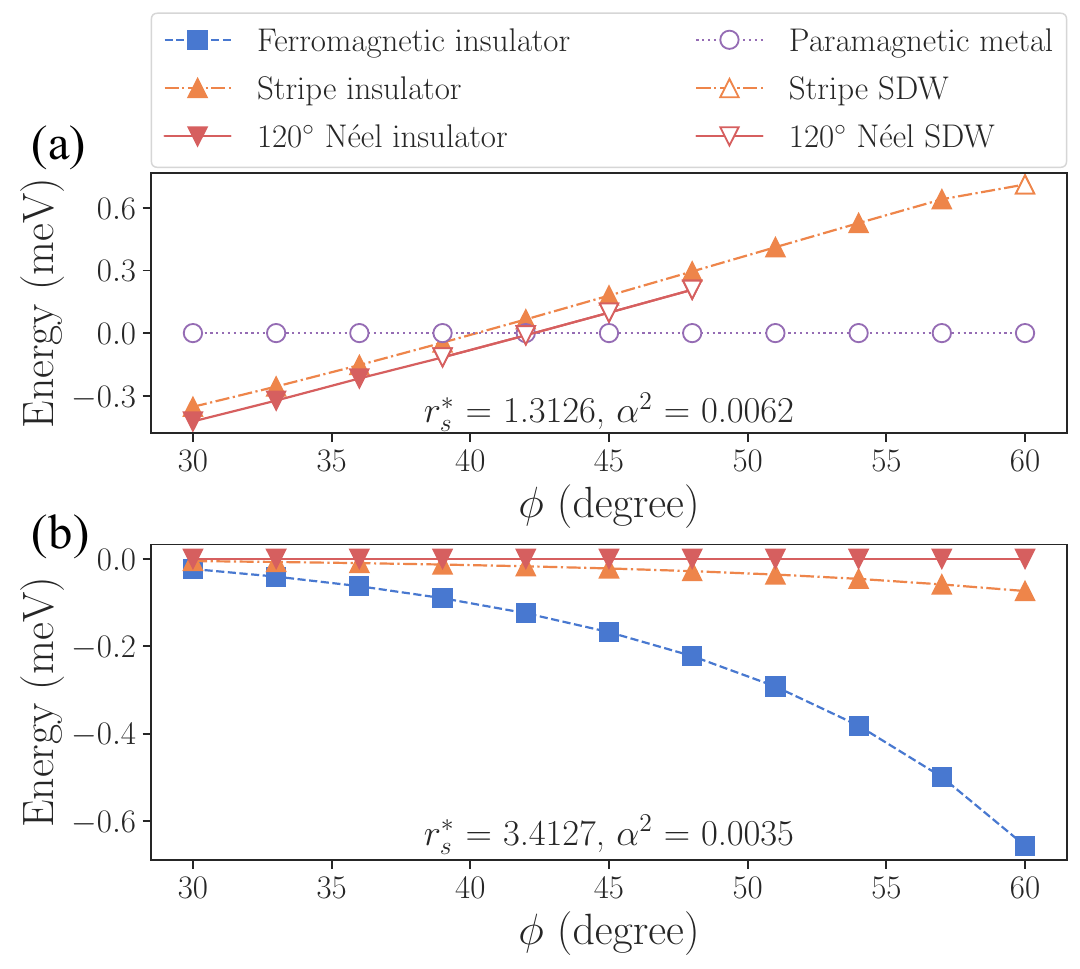}
    \caption{The parameters $(r_s^*, \alpha^2)$ are chosen to be close to (a) the metal-insulator transition and (b) the antiferromagnet-ferromagnet transition at $\phi=26\degree$. We change $\phi$ from $30\degree$ (triangular lattice) to $60\degree$ (honeycomb lattice). (a) Energy density is plotted relative to the paramagnetic states. An insulator-to-metal transition is observed as we increase $\phi$. (b)Energy density is plotted relative to 120$\degree$ N\'eel insulators. We don't see any
    phase transitions in this case.}
    \label{fig:phi}
\end{figure}

From the arguments in Sec.~\ref{sec:hf}, we anticipate the physical effect of changing a triangular lattice to a honeycomb lattice is (approximately) equivalent to increasing the effective Wannier function width $\alpha$.  For this reason we expect the
metal-insulator phase boundary to move towards larger $r_s^*$ as $\phi \to 60\degree$, 
while the antiferromagnet-ferromagnet transition boundary moves towards smaller $r_s^*$.  
For a $r_s^*-\alpha^2$ point on the insulating side of the metal-insulator phase boundary,
we show in Fig.~\ref{fig:phi}(a), that the lowest energy state changes from a noncollinear magnetic 
insulator to a 120$\degree$ semimetallic SDW, and finally to a paramagnetic metal as $\phi \to 60 \degree$.
In contrast, in Fig.~\ref{fig:phi}(b), no phase transition is observed. The ferromagnetic ground state becomes more and more stable as the 
antiferromagnet-ferromagnet phase boundary moves away towards smaller $r_s^*$.  

Last but not least, we examine the possibility of non-trivial band topology in the magnetically ordered states. 
For ferromagnetic and stripe states, it's straightforward to show that the spin-projected orbital Hamiltonians are time-reversal 
invariant.  These states therefore cannot have non-zero Chern numbers. 
The 120$\degree$ non-collinear states also cannot be topological,
since they can be continuously tuned, via intermediate umbrella states, to ferromagnetic states
without closing the gap between occupied and empty states.  More generally any magnetic insulator
that is close to a classical spin state with specific spin-orientations on specific sites cannot have 
a total band Chern number that is non-zero, even if not co-planar.  The quantum anomalous Hall effect
requires itineracy in this sense. The absence of non-trivial band topology is consistent with 
the approach used to approximate the phase boundary analytically. since only in this case are the Wannier functions 
exponentially localized allowing a Gaussian be a good approximation.

Next we make a stronger claim, namely that the intrinsic anomalous Hall conductance is required to be zero even at finite doping 
whenever the magnetic state is co-planar.  
One feature of the heterobilayer continuum Hamiltonian, the lack of inversion symmetry, except at honeycomb values of $\phi$, can lead
to ground states with non-zero momentum-space Berry curvature $\Omega(\bm{k})$ \cite{xiaoBerryrmp2010}. 
In the 120$\degree$ N\'eel state, for example \cite{HuaChenPhysRevLett.112.017205}. 
Because the TRS-breaking 120$\degree$ state has  
$\cal{T}'$ $\equiv$ $\cal{T}$ $\exp (i \pi \b{S}^{\perp})$ symmetry, where $\b{S}^{\perp}$ is the spin operator perpendicular to the 120$\degree$ ordering plane, it follows that  
$\Omega(\bm{k})=-\Omega(-\bm{k})$ and that the topological Chern index obtained by integrating the Berry curvature over the Brillouin-zone vanishes. It's worth noting that this new composite anti-unitary symmetry $\cal{T}'$ squares to $+1$, hence does not imply Kramers degeneracy. (See Fig.~\ref{fig:hfbands} for band structures.) Therefore $\cal{T}'$ is an effective spinless TRS, ensuring that the band structure satisfies 
$E_n(k) = E_n(-k)$. This gives the stronger constraint that anomalous Hall conductivity is zero for all doped system.   

While it's certainly possible to measure this non-zero Berry curvature through a non-linear Hall effect \cite{SodemannNLHE2015}, it's hard to distinguish the ferromagnetic and the 120$\degree$ order, since both share the same qualitative Berry curvature properties.
We propose an alternate strategy to identify the 120$\degree$ state that exploits the proximity of non-collinear 
umbrella \cite{starykh2015unusual} states in which all spins are tilted toward the direction of an applied magnetic field,
breaking $\cal{T}'$. Unlike collinear states, coplanar states can evolve into non-coplanar states under a Zeeman field $\b B$.
$\cal{T}'$ symmetry can only be broken by a non-coplanar magnetic structure. 
It's known, however, coplanar configurations are always favored by quantum fluctuations in a isotropic 
Heisenberg triangular lattice model system under a Zeeman field \cite{chubukov2017j1j2B,farnell2019non}. 
Our proposal therefore relies on anisotropies present in more realistic continuum model beyond the simplest approximation taken by Eq.~\eqref{eq:model}.  If the anisotropy is not strong enough to realize the non-coplanar state, 
the 120$\degree$ coplanar state can also be measured by its distinctive field dependent magnetization curve $\b M(\b B)$. 
Quantum fluctuations favors the collinear UUD state, among the many competetive coplanar states \cite{chubukov2017j1j2B},
at a $\b B$-field of around one-third the saturation value, where $\b M(\b B)$ shows a wide plateau \cite{starykh2015unusual}.
These properties motivate future studies aimed at achieving a full understanding of magnetic anisotropies in 
triangular lattice TMD moir\'e materials.

\section{\label{sec:outlook}Discussion and outlook}

In this paper we have examined the phase diagram of moir\'e Hubbard model systems for the special case of half-filling
of the lowest energy band.  At this filling factor interaction-induced insulating states are normally identified as Mott insulators.
The moir\'e band Hamiltonian \cite{fengchengHubbard} depends on a semiconductor effective mass $m^*$,
the moir\'e potential modulation strength $V_M$, the moir\'e lattice constant $a_M$, 
and in addition on a moir\'e potential shape parameter $\phi$ that interpolates between triangular and honeycomb lattice cases.  The interaction term
is sensitive to screening by polarizable backgrounds (including but not necessarily limited to \cite{liu2021tuning} 
screening by the surrounding dielectric), which we 
characterize collectively by an effective inverse dielectric constant $\epsilon^{-1}$.
At fixed band filling, the model parameters can be collapsed to the shape parameter $\phi$ and 
two dimensionless coupling constant ratios, $r_s^*$ and $\alpha^2$,
chosen with the goal of minimizing the dependence of the phase diagram on $\phi$.  
$r_s^*$ is the standard
electron gas density parameter and $\alpha^2$ is proportional to the fraction of the unit cell area 
occupied by the model's Wannier orbital.  
We find a phase diagrams with two prominent transitions, an expected Mott transition between metallic and 
insulating states, and an unexpected transition between antiferromagnetic and ferromagnetic insulating states.
We predict that the metal-insulator transition occurs along a line of nearly constant $U/t_1\sim 15.1$, where $U$ is the 
on-site Hubbard interaction and $t_1$ is the triangular lattice near-neighbor hopping parameter.  The value of this ratio 
on the metal-insulator transition line is comparable to values obtained in numerical studies of simple on-site-interaction 
triangular lattice Hubbard models \cite{PhysRevX.10.021042,becca2020Utratio}. 

The metal-insulator transition line can be crossed by changing the electron density parameter $r_s^*$ by 
changing twist angle, by engineering the depth of the modulating moir\'e potential via suitable choice of materials,
or {\em in situ} by tuning gate voltages \cite{mak2021continuousMIT,dean2021quantumcritical} or applying pressure \cite{Yankowitz1059}.
On the insulating side of the metal-insulator transition we find the 120$\degree$ three-sublattice antiferromagnet expected on triangular lattices
with antiferromagnetic interactions between spins.  Within the Hartree-Fock approximation, we find a narrow band of intermediate semimetallic 
states that maintain the 120$\degree$ semimetallic SDW order of the insulating state.  The phase transition between the SDW state 
and the strongly metallic state is first order.

The SDW phase that appears in our calculations provides one possible explanation for the
complex crossover between insulating and metallic states seen in recent experiments  \cite{mak2021continuousMIT,dean2021quantumcritical},
which hint at an intermediate state with a small but finite zero-temperature
conductivity \footnote{A resistance jump also occurs at the critical point in electron-fractionalization \cite{senthil2008mit,xu2021metal} metal-insulator transition scenarios, but 
for completely different reasons.}.

In closing we comment that in this MS we focused on the simplest case in which electronic states are formed from a single 
microscopic band, and therefore described in a continuum model by two-component spinors.  In the case of 
TMD homobilayers \cite{PhysRevLett.122.086402,pan2020band,zang2021hartree,devakul2021magic}, and in the case of heterobilayers modified by suitable large gate electric fields \cite{zhang2021spin}, low-energy 
bands are present in both layers, yielding low-energy models with four component spinors that capture
both spin and layer degrees of freedom and opening up new opportunities to establish topologically non-trivial states.  Indeed recent heterobilayer experiments  \cite{li2021quantum} find transitions to states with spontaneous valley polarization and 
anomalous Hall effects in TMDs.  We have also limited our attention to one electron per moir\'e period.  
Doping way from this limit, enriches the physics even more and is thought to lead to superconductivity in some cases.
Finally, we have neglected disorder, which might be relevant experimentally, as suggested by the temperature-dependent 
resistances measured in Ref.~\cite{mak2021continuousMIT}, which exhibit 
bumps on the metalic side of the metal-insulator transition similar to the ones seen in Si-MOSFETs \cite{finkelstein2001silicon}.
The situation studied in this MS is explores only one simplest limit of the rich physics 
that remains to be explored in TMD moir\'e materials.  

\begin{acknowledgments}
We acknowledge helpful discussions with Nicol\'as Morales-Dur\'an, Pawel Potasz, Ajesh Kumar and Jihang Zhu. 
This work was supported by the U.S. Department of Energy, Office of Science, Basic Energy Sciences, under Award DE-FG02-02ER45958. The authors acknowledge the Texas Advanced Computing Center (TACC) at The University of Texas at Austin for providing HPC resources that have contributed to the research results reported within this paper.
\end{acknowledgments}

\appendix

\section{\label{app:cutoff}More details on the self-consistent Hartree-Fock calculations}

\begin{figure}[!t]
    \centering
    \includegraphics[width=0.45\textwidth]{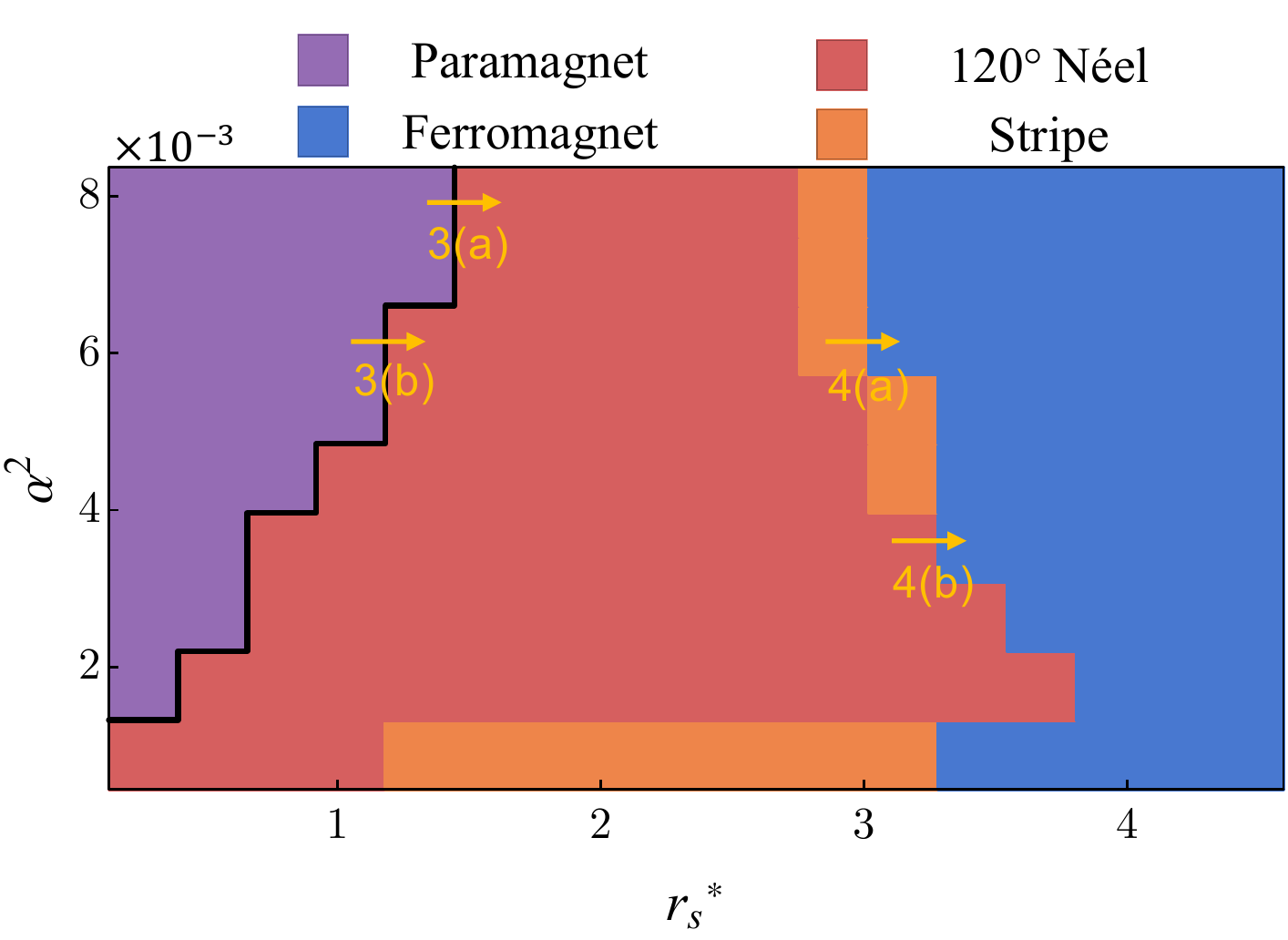}
    \caption{Pixelated phase diagram for triangular lattice ($\phi=26\degree$) moir\'e materials within Hartree-Fock approximation.}
    \label{fig:pixel}
\end{figure}

To make the continuum model feasible for a numerical calculation, it's necessary to impose both upper- and lower-cutoffs in the momentum space. In the atomic-insulator limit, the real-space Wannier function is well approximated by Gaussian in a harmonic potential, as argued in the main text. It's then clear that the relevant scale for the upper-cutoff can be obtained by comparing the momentum-space Wannier function width with the moir\'e Brillouin zone size:
\begin{align}
    \frac{1}{a_W}\bigg/|\vec{b}| = \sqrt{\frac{m^*\omega}{\hbar}}\bigg/|\vec{b}| = \frac{\sqrt{3}}{4\pi}\(\frac{\beta m^*V_M}{\hbar^2}\)^{1/4}\sqrt{a_M}.
\end{align} 
That is to say a larger upper-cutoff may be required for stronger interactions. The resulting dimensionless number is of $\mathcal{O}(1)$ in our calculation, and we keep the number of momentum shells in the continuum model such that more shells don't change the Hartree-Fock energies up to the convergence accuracy. 

As for the lower-cutoff, we note that energy density for a periodic system with Coulomb interactions suffers from a finite-size correction, which depends not only on the total number of mesh points $N$ in the moir\'e Brillouin zone but also on the geometrical detail of the mesh. We correct the finite-size effect by choosing the geometry such that it's compatible with all the pertinent magnetic moir\'e Brillouin zones, and meshing the moir\'e Brillouin zone always in the same way. In other words, we hold the real-space sample size $N$ fixed for all calculations, which is $108$ in our phase diagram calculations. Since the leading finite-size corrections to energy are now state independent, they can be 
eliminated by evaluating energy differences between states at fixed system size.
We do not actually extrapolate the energies toward the thermodynamic limit since the extrapolation itself incurs further uncertainty.

The actual phase diagram from the self-consistent Hartree-Fock calculation is shown in Fig.~\ref{fig:pixel}. Fig.~\ref{fig:phaseD} in the main text is obtained by fitting the analytical form of the phase boundary to Fig.~\ref{fig:pixel}. In the limit of $\alpha\rightarrow 0$, we observe different spin configurations converge to the same energy due to the vanishing of the exchange energies $J$, which we do not show in the main text. 

To show the Hartree-Fock band structures in Fig.~\ref{fig:hfbands}(b) with a clear small Fermi surface, we increase the system size $N$ to $432$. A caveat here is that when we consider a specific magnetic ordered insulating state, the gap size is dependent on the system size because the interaction $U\propto -1/\sqrt{N}$. Hence there is a slight mismatch in parameters and phases between Fig.~\ref{fig:hfbands} and \ref{fig:mit}. This finite-size correction to the gap size is small when $N$ is large, so the phase boundary of the 120$\degree$ semimetallic SDW state is still relatively accurate in Fig.~\ref{fig:phaseD}.  

\section{\label{app:tU}Perturbative effects of ``nonlocal'' interactions}
In this section, we illustrate the induced spin-spin interactions by $V_{ah}$ and $V_x$ in the presence of a large-$U$ Hubbard interaction. For simplicity, we always consider the half-filled case and zero-hopping limit. The ground state lies in the no doubly-occupied site sector as in a usual Hubbard model. It can be readily seen that $V_{ah}$ perturbs the ground state out of this sector, so the lowest order contribution is of order $\mathcal{O}(|V_{ah}|^2/U)$. In spirit of the $t/U$ expansion \cite{allanPhysRevB.37.9753}, the leading order terms are
\begin{widetext}
\begin{align}
    -U^{-1}\sum_{\sigma_1,\sigma_2}&\[\braket{2\sigma_2,1\bar{\sigma}_2|V_C|1\sigma_2,1\bar{\sigma}_2}\braket{1\sigma_1,1\bar{\sigma}_1|V_C|1\sigma_1,2\bar{\sigma}_1}c^{\dag}_{2\sigma_2}c^{\dag}_{1\bar{\sigma}_2}c_{1\bar{\sigma}_2}c_{1\sigma_2}c^{\dag}_{1\sigma_1}c^{\dag}_{1\bar{\sigma}_1}c_{2\bar{\sigma}_1}c_{1\sigma_1}+\right. \nonumber\\
    & \left. \braket{1\sigma_2,2\bar{\sigma}_2|V_C|1\sigma_2,1\bar{\sigma}_2}\braket{1\sigma_1,1\bar{\sigma}_1|V_C|1\sigma_1,2\bar{\sigma}_1}c^{\dag}_{1\sigma_2}c^{\dag}_{2\bar{\sigma}_2}c_{1\bar{\sigma}_2}c_{1\sigma_2}c^{\dag}_{1\sigma_1}c^{\dag}_{1\bar{\sigma}_1}c_{2\bar{\sigma}_1}c_{1\sigma_1}\]+h.c.,\label{eq:vah}
\end{align}
\end{widetext}
where $1$ and $2$ label nearest neighbors.
The fermionic interactions of the two terms in the square brackets are actually related by a relabeling symmetry: the second term $(\sigma_2\rightarrow\bar{\sigma}_2)$ = the first term. The first term in Eq.~\eqref{eq:vah} can be simplified, using $\sum_{\sigma}n_{i\sigma} = 1$, to $-|V_{ah}|^2U^{-1}\(-\sum_{\sigma}c^{\dag}_{1\sigma}c_{1\bar{\sigma}}c^{\dag}_{2\bar{\sigma}}c_{2{\sigma}}+\sum_{\sigma}n_{1\sigma}n_{2\bar{\sigma}}\) = -|V_{ah}|^2U^{-1} \(1-\bm{\sigma}_1\cdot\bm{\sigma}_2\)/2$, where $\bm{\sigma}_i = c^{\dag}_{i\alpha}\bm{\sigma}_{\alpha\beta}c_{i\beta}$ and $|V_{ah}|^2 = \braket{2,1|V_C|1,1}\braket{1,1|V_C|1,2}$. Hence it gives rise to the same type of contribution as the normal hopping term in Hubbard model, i.e. antiferromagnetic coupling. 

Now we turn to consider the effects of $V_x$. $V_x$ actually leaves the number of doubly occupied sites invariant. So the lowest order contribution is just itself:
\begin{align}
    &\sum_{\sigma_1,\sigma_2}\braket{2\sigma_1,1\sigma_2|V_C|1\sigma_1,2\sigma_2}c^{\dag}_{2\sigma_1}c^{\dag}_{1\sigma_2}c_{2\sigma_2}c_{1\sigma_1}\\
    =& \braket{2,1|V_C|1,2}\sum_{\sigma_1,\sigma_2}c^{\dag}_{2\sigma_1}c^{\dag}_{1\sigma_2}c_{2\sigma_2}c_{1\sigma_1}\\
    =& -\frac{\braket{2,1|V_C|1,2}}{2}\(1+\bm{\sigma}_1\cdot\bm{\sigma}_2\),
\end{align}
which favors ferromagnetic spin configuration. 

\bibliography{bib}

\end{document}